# Transformation optics in orthogonal coordinates

Huanyang Chen[a, b, *]

[a] *Department of Physics, The Hong Kong University of Science and Technology,*

*Clear Water Bay, Hong Kong, China*

[b] *Institute of Theoretical Physics, Shanghai Jiao Tong University, Shanghai 200240,*

*China*

* Corresponding author at: Department of Physics, The Hong Kong University of Science and Technology, Clear Water Bay, Hong Kong, China.

Tel.: +852 2358 7981; fax: +852 2358 1652.

*E-mail address:* kenyon@ust.hk (Huanyang Chen).

**Abstract:**

The author proposes the methodology of transformation optics in orthogonal coordinates to obtain the material parameters of the transformation media from the mapping in orthogonal coordinates. Several examples are given to show the applications of such a methodology by using the full-wave simulations.

## I. Introduction

The transformation optics [1, 2] have provided a general method to design all kinds of devices with novel properties, such as the invisibility cloak [1], concentrator [3], rotation cloak [4], electromagnetic wormhole [5], and impedance-matched hyperlens [6], *etc*. The invisibility cloak is the most attracting topics. Cloaks with all kinds of shapes have been designed [3, 7-10]. Specifically, a cylindrical cloak with arbitrary cross section has been given in [10], which has been extended to design the perfect electrical conductor (PEC) reshaper [11] and other devices with arbitrary cross sections. In this paper, I will go further to rewrite the transformation media equations from the Cartesian coordinates to orthogonal coordinates. Several examples will be demonstrated to show the applications of such transformation optics. We note that the technique here is not a new theory but to rewrite the formula in a convenient form and to show an easy way to obtain the material parameters which would be very laborious from original transformation optics in Cartesian coordinate. The reader can also refer a more general technique proposed by Leonhardt and Philbin to study the "general relativity in electrical engineering" [12, 13], where they started from Maxwell's equations in curved coordinates and used standard results from differential geometry. The technique here focuses to write out the material parameters in a very explicit way for the future study by means of the numerical finite element methods. We should also differentiate the Cartesian coordinates from the general curvilinear orthogonal coordinates because the Cartesian coordinate itself is orthogonal. When we talk about the orthogonal coordinates in this paper, it usually means other orthogonal coordinates but not the Cartesian coordinates. The paper is organized as follows. Section I is the introduction part. Section II is the theory part for transformation optics in orthogonal coordinates. Section III and IV are two applications of the above transformation optics. Section V is a summary of this paper.

## II. Transformation optics in orthogonal coordinate.

Let's start from the transformation media equations in Cartesian coordinate [14, 15],

$$\ddot{\varepsilon}_{Cartesian}{}^{i'j'} = J_i^{i'} J_j^{j'} \ddot{\varepsilon}_{Cartesian}{}^{ij} / \det(J_i^{i'}),$$
$$\ddot{\mu}_{Cartesian}{}^{i'j'} = J_i^{i'} J_j^{j'} \ddot{\mu}_{Cartesian}{}^{ij} / \det(J_i^{i'}),$$
(1)

where $J_i^{i'} = \dfrac{\partial x^{i'}}{\partial x^i}$ is the Jacobian transformation matrix in Cartesian coordinate, $\ddot{\varepsilon}_{Cartesian}{}^{ij}$ and $\ddot{\mu}_{Cartesian}{}^{ij}$ are the components of the permittivity and permeability tensors in the original space in Cartesian coordinate, $\ddot{\varepsilon}_{Cartesian}{}^{i'j'}$ and $\ddot{\mu}_{Cartesian}{}^{i'j'}$ are the components of the equivalent permittivity and permeability tensors of the transformation media in Cartesian coordinate to obtain the transformed space [14, 15].

Now we rewrite the components of the permittivity and permeability tensors in the original space in Cartesian coordinate in the form of the ones in the original space in an orthogonal coordinate,

$$\ddot{\varepsilon}_{Cartesian}{}^{ij} = R_{0k}^{i} R_{0l}^{j} \ddot{\varepsilon}_{Orthogonal}{}^{kl} / \det(R_{0k}^{i}),$$
$$\ddot{\mu}_{Cartesian}{}^{ij} = R_{0k}^{i} R_{0l}^{j} \ddot{\mu}_{Orthogonal}{}^{kl} / \det(R_{0k}^{i}),$$
(2)

where $R_{0k}^{i} = \frac{\partial x^i}{h_{u^k} \partial u^k}$ is the rotation matrix between the orthogonal coordinate and the Cartesian coordinate. The line element in Cartesian coordinate in the original space is $ds^2 = dx^2 + dy^2 + dz^2 = \delta_{ij} dx^i dx^j$, while in the orthogonal coordinate in the original space, it is

$$ds^2 = h_u(u,v,w)^2 du^2 + h_v(u,v,w)^2 dv^2 + h_w(u,v,w)^2 dw^2$$
$$= h_u^2 du^2 + h_v^2 dv^2 + h_w^2 dw^2 = \delta_{ij} h_{u^i} h_{u^j} du^i du^j.$$
(3)

Substitute Eq. (2) in to Eq. (1), we have,

$$\ddot{\varepsilon}_{Cartesian}{}^{i'j'} = J_i^{i'} J_j^{j'} R_{0k}^{i} R_{0l}^{j} \ddot{\varepsilon}_{Orthogonal}{}^{kl} / \det(J_i^{i'}) / \det(R_{0k}^{i})$$
$$= Q_k^{i'} Q_l^{j'} \ddot{\varepsilon}_{Orthogonal}{}^{kl} / \det(Q_k^{i'}),$$
$$\ddot{\mu}_{Cartesian}{}^{i'j'} = J_i^{i'} J_j^{j'} R_{0k}^{i} R_{0l}^{j} \ddot{\mu}_{Orthogonal}{}^{kl} / \det(J_i^{i'}) / \det(R_{0k}^{i})$$
$$= Q_k^{i'} Q_l^{j'} \ddot{\mu}_{Orthogonal}{}^{kl} / \det(Q_k^{i'}),$$
(4)

with $Q_k^{i'} = J_i^{i'} R_{0k}^{i} = \frac{\partial x^{i'}}{\partial x^i} \frac{\partial x^i}{h_{u^k} \partial u^k} = \frac{\partial x^{i'}}{h_{u^k} \partial u^k}$.

We rewrite the components of the equivalent permittivity and permeability tensors of the transformation media from the Cartesian coordinate to the orthogonal coordinate,

$$\ddot{\varepsilon}_{Orthogonal}{}^{k'l'} = R^{-1k'}{}_{i'}R^{-1l'}{}_{j'}\ddot{\varepsilon}_{Cartesian}{}^{i'j'} / \det(R^{-1k'}{}_{i'}),$$
$$\ddot{\mu}_{Orthogonal}{}^{k'l'} = R^{-1k'}{}_{i'}R^{-1l'}{}_{j'}\ddot{\mu}_{Cartesian}{}^{i'j'} / \det(R^{-1k'}{}_{i'}),$$
(5)

with the rotation matrix $R^{-1k'}{}_{i'} = \dfrac{h_{u^{k'}}\partial u^{k'}}{\partial x^{i'}}$. The line element in Cartesian coordinate in the transformed space is $ds'^2 = dx'^2 + dy'^2 + dz'^2 = \delta_{i'j'}dx^{i'}dx^{j'}$, while in the orthogonal coordinate in the transformed space, it is

$$ds'^2 = h_{u'}(u',v',w')^2 du'^2 + h_{v'}(u',v',w')^2 dv'^2 + h_{w'}(u',v',w')^2 dw'^2$$
$$= h_{u'}^2 du'^2 + h_{v'}^2 dv'^2 + h_{w'}^2 dw'^2 = \delta_{i'j'}h_{u^{i'}}h_{u^{j'}}du^{i'}du^{j'}.$$
(6)

Substitute Eq. (4) in to Eq. (5),

$$\ddot{\varepsilon}_{Orthogonal}{}^{k'l'} = R^{-1k'}{}_{i'}R^{-1l'}{}_{j'}Q^{i'}_{k}Q^{j'}_{l}\ddot{\varepsilon}_{Orthogonal}{}^{kl} / \det(Q^{i'}_{k}) / \det(R^{-1k'}{}_{i'})$$
$$= T^{k'}_{k}T^{l'}_{l}\ddot{\varepsilon}_{Orthogonal}{}^{kl} / \det(T^{k'}_{k}),$$
$$\ddot{\mu}_{Orthogonal}{}^{k'l'} = R^{-1k'}{}_{i'}R^{-1l'}{}_{j'}Q^{i'}_{k}Q^{j'}_{l}\ddot{\mu}_{Orthogonal}{}^{kl} / \det(Q^{i'}_{k}) / \det(R^{-1k'}{}_{i'})$$
$$= T^{k'}_{k}T^{l'}_{l}\ddot{\mu}_{Orthogonal}{}^{kl} / \det(T^{k'}_{k}),$$
(7)

with $T^{k'}_{k} = R^{-1k'}{}_{i'}Q^{i'}_{k} = \dfrac{h_{u^{k'}}\partial u^{k'}}{\partial x^{i'}}\dfrac{\partial x^{i'}}{h_{u^{k}}\partial u^{k}} = \dfrac{h_{u^{k'}}\partial u^{k'}}{h_{u^{k}}\partial u^{k}}$. Equation (7) is the transformation media equation in the orthogonal coordinate. More explicitly, suppose the mapping is written as $(u,v,w) \Leftrightarrow (u',v',w')$,

$$\begin{cases} u' = u'(u,v,w), \\ v' = v'(u,v,w), \\ w' = w'(u,v,w), \end{cases} \text{or} \begin{cases} u = u(u',v',w'), \\ v = v(u',v',w'), \\ w = w(u',v',w'), \end{cases}$$
(8)

the Jacobian transformation matrix in the orthogonal coordinate is,

$$T = \begin{bmatrix} \dfrac{h_{u'}}{h_u}\dfrac{\partial u'}{\partial u} & \dfrac{h_{u'}}{h_v}\dfrac{\partial u'}{\partial v} & \dfrac{h_{u'}}{h_w}\dfrac{\partial u'}{\partial w} \\ \dfrac{h_{v'}}{h_u}\dfrac{\partial v'}{\partial u} & \dfrac{h_{v'}}{h_v}\dfrac{\partial v'}{\partial v} & \dfrac{h_{v'}}{h_w}\dfrac{\partial v'}{\partial w} \\ \dfrac{h_{w'}}{h_u}\dfrac{\partial w'}{\partial u} & \dfrac{h_{w'}}{h_v}\dfrac{\partial w'}{\partial v} & \dfrac{h_{w'}}{h_w}\dfrac{\partial w'}{\partial w} \end{bmatrix},$$
(9)

the transformation media equation is,

$$\ddot{\varepsilon}_{Orthogonal}' = T\ddot{\varepsilon}_{Orthogonal}T^T / \det(T),$$
$$\ddot{\mu}_{Orthogonal}' = T\ddot{\mu}_{Orthogonal}T^T / \det(T),$$
(10)

in the form of $(u',v',w')$.

To use the numerical finite element methods (for instance, the COMSOL Multiphysics finite element-based electromagnetics solver), we should obtain the explicit form of the required parameters in the form of $(x',y',z')$ [16]. The rotation matrix is,

$$R = \begin{bmatrix} \frac{1}{h_{u'}}\frac{\partial x'}{\partial u'} & \frac{1}{h_{v'}}\frac{\partial x'}{\partial v'} & \frac{1}{h_{w'}}\frac{\partial x'}{\partial w'} \\ \frac{1}{h_{u'}}\frac{\partial y'}{\partial u'} & \frac{1}{h_{v'}}\frac{\partial y'}{\partial v'} & \frac{1}{h_{w'}}\frac{\partial y'}{\partial w'} \\ \frac{1}{h_{u'}}\frac{\partial z'}{\partial u'} & \frac{1}{h_{v'}}\frac{\partial z'}{\partial v'} & \frac{1}{h_{w'}}\frac{\partial z'}{\partial w'} \end{bmatrix},$$
(11)

Finally, the material parameters are,

$$\ddot{\varepsilon}_{Cartesian}' = RT\ddot{\varepsilon}_{Orthogonal}T^T R^T / \det(T) / \det(R),$$
$$\ddot{\mu}_{Cartesian}' = RT\ddot{\mu}_{Orthogonal}T^T R^T / \det(T) / \det(R).$$
(12)

One can find more detailed procedures on using Eq. (12) in the following sections.

## III. An application of the transformation optics in orthogonal coordinate

Now we come to an application of the above transformation optics. Let's suppose,

$$\ddot{\varepsilon}_{Orthogonal} = \begin{bmatrix} \varepsilon_u & 0 & 0 \\ 0 & \varepsilon_v & 0 \\ 0 & 0 & \varepsilon_w \end{bmatrix},$$
$$\ddot{\mu}_{Orthogonal} = \begin{bmatrix} \mu_u & 0 & 0 \\ 0 & \mu_v & 0 \\ 0 & 0 & \mu_w \end{bmatrix},$$
(13)

the mapping is $(u,v,w) \Leftrightarrow (u',v',w')$ with,

$$\begin{cases} u' = u'(u), \\ v' = v'(v), \\ w' = w'(w), \end{cases} \text{ or } \begin{cases} u = u(u'), \\ v = v(v'), \\ w = w(w'), \end{cases} \tag{14}$$

the Jacobian transformation matrix in the orthogonal coordinate is,

$$T = \begin{bmatrix} \frac{h_{u'}}{h_u}\frac{\partial u'}{\partial u} & 0 & 0 \\ 0 & \frac{h_{v'}}{h_v}\frac{\partial v'}{\partial v} & 0 \\ 0 & 0 & \frac{h_{w'}}{h_w}\frac{\partial w'}{\partial w} \end{bmatrix} = \begin{bmatrix} \frac{1}{Q_u} & 0 & 0 \\ 0 & \frac{1}{Q_v} & 0 \\ 0 & 0 & \frac{1}{Q_w} \end{bmatrix}, \tag{15}$$

from Eq. (7) or Eq. (10), the equivalent permittivity and permeability tensors of the transformation media are,

$$\ddot{\varepsilon}_{Orthogonal}' = \begin{bmatrix} \varepsilon_u \frac{Q_u Q_v Q_w}{Q_u^2} & 0 & 0 \\ 0 & \varepsilon_v \frac{Q_u Q_v Q_w}{Q_v^2} & 0 \\ 0 & 0 & \varepsilon_w \frac{Q_u Q_v Q_w}{Q_w^2} \end{bmatrix},$$

$$\ddot{\mu}_{Orthogonal}' = \begin{bmatrix} \mu_u \frac{Q_u Q_v Q_w}{Q_u^2} & 0 & 0 \\ 0 & \mu_v \frac{Q_u Q_v Q_w}{Q_v^2} & 0 \\ 0 & 0 & \mu_w \frac{Q_u Q_v Q_w}{Q_w^2} \end{bmatrix}, \tag{16}$$

which are exactly the original results obtained in Ref. [1].

We could also start from the method in [12] to obtain the same results. For example, if $\ddot{\varepsilon}_{Orthogonal} = \ddot{\mu}_{Orthogonal} = diag\{1,1,1\}$, the corresponding metric tensor here is,

$$g_{ij} = diag\{[\frac{h_u}{h_{u'}}\frac{\partial u}{\partial u'}]^2, [\frac{h_v}{h_{v'}}\frac{\partial v}{\partial v'}]^2, [\frac{h_w}{h_{w'}}\frac{\partial w}{\partial w'}]^2\} = diag\{Q_u^2, Q_v^2, Q_w^2\},$$

(17)

so that $\sqrt{g} = Q_u Q_v Q_w$. Then with the help of $\varepsilon^{ij} = \mu^{ij} = \sqrt{g} g^{ij}$, we have,

$$\varepsilon^{ij} = \mu^{ij} = diag\{\frac{Q_u Q_v Q_w}{Q_u^2}, \frac{Q_u Q_v Q_w}{Q_v^2}, \frac{Q_u Q_v Q_w}{Q_w^2}\}, \quad (18)$$

The method in [12] can be applied to produce the results in Sec. II as well in the case of vacuum in the original space. However, if there are some materials in the original space, further modifications should be addressed.

(a) Transformation media in the elliptic cylindrical coordinate

With the above simplification, we first come to a specific example in the elliptic cylindrical coordinate. The elliptic cylindrical coordinate $(u, v, z)$ and the Cartesian coordinate have the following relationship,

$x = f \cosh u \cos v,$

$y = f \sinh u \sin v,$ (19)

$z = z,$

where the distance of the two focuses is $2f$, and its line element is,

$$ds^2 = a^2(\sinh^2 u + \sin^2 v)du^2 + a^2(\sinh^2 u + \sin^2 v)dv^2 + dz^2 \\ = h_u^2 du^2 + h_v^2 dv^2 + dz^2 = \delta_{ij} h_{u^i} h_{u^j} du^i du^j. \quad (20)$$

Suppose the mapping is: $u = u(u')$, $v = v'$, and $\vec{\vec{\varepsilon}}_{Orthogonal} = \vec{\vec{\mu}}_{Orthogonal} = \vec{\vec{I}}$, from Eq. (16),

$\varepsilon_{u'} = \mu_{u'} = \dfrac{1}{du/du'},$

$\varepsilon_{v'} = \mu_{v'} = du/du',$ (21)

$\varepsilon_{z'} = \mu_{z'} = \dfrac{du}{du'} \dfrac{\cosh^2 u(u') - \cos^2 v'}{\cosh^2 u' - \cos^2 v'},$

with the help of its rotation matrix, we can obtain the permittivity and permeability tensors of the transformation media in Cartesian coordinate,

$$\varepsilon_{x'x'} = \varepsilon_{u'} \frac{\sinh^2 u' \cos^2 v'}{\cosh^2 u' - \cos^2 v'} + \varepsilon_{v'} \frac{\cosh^2 u' \sin^2 v'}{\cosh^2 u' - \cos^2 v'},$$

$$\varepsilon_{x'y'} = \varepsilon_{y'x'} = (\varepsilon_{u'} - \varepsilon_{v'}) \frac{\cosh u' \cos v' \sinh u' \sin v'}{\cosh^2 u' - \cos^2 v'}, \qquad (22)$$

$$\varepsilon_{y'y'} = \varepsilon_{u'} \frac{\cosh^2 u' \sin^2 v'}{\cosh^2 u' - \cos^2 v'} + \varepsilon_{v'} \frac{\sinh^2 u' \cos^2 v'}{\cosh^2 u' - \cos^2 v'},$$

with $\vec{\vec{\mu}} = \vec{\vec{\varepsilon}}$ completing the material tensor description.

We consider the following mapping from an elliptic cylinder region ($0 < u < u_2$) to an elliptic cylindrical shell ($u_1 < u' < u_2$),

$$u' = u_1 + u(u_2 - u_1)/u_2, \text{ or } u = \frac{u' - u_1}{u_2 - u_1} u_2. \qquad (23)$$

where a line ($u = 0$) is mapped into an ellipse ($u' = u_1$). After this mapping, a kind of transformation media could be obtained, which shall be called the "elliptic cylindrical cloak" [7]. The elliptic cylindrical cloak can reduce a PEC elliptic cylinder in to a PEC plate in the view of transformation optics. From Eq. (21), the material parameters are,

$$\mu_{u'} = \frac{u_2 - u_1}{u_2},$$

$$\mu_{v'} = \frac{u_2}{u_2 - u_1}, \qquad (24)$$

$$\varepsilon_{z'} = \frac{u_2}{u_2 - u_1} \frac{\cosh^2(\frac{u' - u_1}{u_2 - u_1} u_2) - \cos^2 v'}{\cosh^2 u' - \cos^2 v'}.$$

Here we only consider the transverse electric (TE) polarization incident waves interacting with the transformation media. One can substitute Eq. (24) into Eq. (22) to obtain the components in the form of $(x', y', z')$.

As a concrete example, we set $u_1 = 1\,m$ and $u_2 = 2\,m$. The distance between two focuses is 0.14 m. The frequency of the incident plane wave is 2 GHz. Fig. 1(a) is the scattering pattern of a PEC plate and the plane wave is incident from left to right (x-direction). The scattering is small in this incident direction. The PEC plate locates at x-axis from -0.07m to 0.07m, the width in y-direction is infinitesimal. Fig. 1(b) is the scattering pattern of the PEC plate and the plane wave is incident from bottom to top (y-direction). The scattering is considerable large in this incident direction. In fig. 1(c), we plot the scattering pattern of a PEC cylinder coating with the elliptic cylindrical cloak, and the plane wave is incident from left to right. The far-field scattering pattern in fig. 1(c) is the same to the one in fig. 1(a). That means the elliptic cylindrical cloak can reduce the scattering of a PEC cylinder in its long-axis direction. In fig. 1(d), we plot the scattering pattern of a PEC cylinder coating with the elliptic cylindrical cloak, and the plane wave is incident from bottom to top. The far-field scattering pattern in fig. 1(d) is the same to the one in fig. 1(b). That means the elliptic cylindrical cloak cannot reduce the scattering of a PEC cylinder in its short-axis direction. The elliptic cylindrical cloak can be treated as the one dimensional cloak in literature [17, 18], which seems to show a way to the broadband cloaks by sacrificing the cross section for the bandwidth [19]. Recently, Leonhardt and Tyc have given a completely different method to obtain a perfect broadband cloak (non-Euclidean cloak) [20] with time delays in the regime of geometrical optics.

## (b) Transformation media in bipolar cylindrical coordinate

Now we come to another example, which is the transformation media in bipolar cylindrical coordinate. The bipolar cylindrical coordinate $(u,v,z)$ and the Cartesian coordinate have the following relationship,

$$x = f\frac{\sinh u}{\cosh u - \cos v},$$
$$y = f\frac{\sin v}{\cosh u - \cos v}, \qquad (25)$$
$$z = z,$$

where the distance of the two focuses is $2f$, and its line element,

$$ds^2 = (\frac{a}{\cosh v - \cos u})^2 du^2 + (\frac{a}{\cosh v - \cos u})^2 dv^2 + dz^2$$
$$= h_u^2 du^2 + h_v^2 dv^2 + dz^2 = \delta_{ij} h_{u^i} h_{u^j} du^i du^j. \qquad (26)$$

Suppose the mapping is: $u = u(u')$, $v = v'$, and $\ddot{\varepsilon}_{Orthogonal} = \ddot{\mu}_{Orthogonal} = \ddot{I}$, from Eq. (16),

$$\varepsilon_{u'} = \mu_{u'} = \frac{1}{du/du'},$$
$$\varepsilon_{v'} = \mu_{v'} = du/du', \qquad (27)$$
$$\varepsilon_{z'} = \mu_{z'} = \frac{du}{du'} \frac{(\cosh u' - \cos v')^2}{(\cosh u(u') - \cos v')^2},$$

with the help of its rotation matrix, we can obtain the permittivity and permeability tensors of the transformation media in Cartesian coordinate,

$$\varepsilon_{x'x'} = \varepsilon_{u'}\frac{(\cosh u'\cos v'-1)^2}{(\cosh u'-\cos v')^2} + \varepsilon_{v'}\frac{\sinh^2 u'\sin^2 v'}{(\cosh u'-\cos v')^2},$$

$$\varepsilon_{x'y'} = \varepsilon_{y'x'} = (\varepsilon_{u'} - \varepsilon_{v'})\frac{\sinh u'\sin v'(\cosh u'\cos v'-1)}{(\cosh u'-\cos v')^2}, \qquad (28)$$

$$\varepsilon_{y'y'} = \varepsilon_{u'} \frac{\sinh^2 u' \sin^2 v'}{(\cosh u' - \cos v')^2} + \varepsilon_{v'} \frac{(\cosh u' \cos v' - 1)^2}{(\cosh u' - \cos v')^2},$$

with $\vec{\vec{\mu}} = \vec{\vec{\varepsilon}}$ completing the material tensor description.

We consider the following mapping from a cylinder region ($u_2 < u < +\infty$) to a non-concentric cylindrical shell ($u_2 < u' < u_1$),

$$u' = u_2(u_2 - u_1)\frac{1}{u} + u_1, \text{ or } u = \frac{u_2(u_2 - u_1)}{u' - u_1}, \tag{29}$$

where a point ($u = +\infty$) is mapped into a circle ($u' = u_1$). After this mapping, a kind of transformation media could be obtained, which we shall call it the "bipolar cylindrical cloak". The bipolar cylindrical cloak is a perfect cloak like the first version of circular cylindrical cloak [16] in the view of transformation optics. From Eq. (27), the material parameters are,

$$\mu_{u'} = \frac{(u' - u_1)^2}{u_2(u_1 - u_2)},$$

$$\mu_{v'} = \frac{u_2(u_1 - u_2)}{(u' - u_1)^2}, \tag{30}$$

$$\varepsilon_{z'} = \frac{u_2(u_1 - u_2)}{(u' - u_1)^2} \frac{(\cosh u' - \cos v')^2}{(\cosh \frac{u_2(u_2 - u_1)}{u' - u_1} - \cos v')^2}.$$

As a concrete example, we set $u_1 = 1\,m$ and $u_2 = 0.5\,m$. The distance between two focuses is 0.14 m. The frequency of the incident plane wave is 2 GHz. Fig. 2(a) and (b) show the scattering pattern of the bipolar cylindrical cloak interacting with an incident plane wave from left to right and from bottom to top respectively. The perfect cloaking effect can be seen visually from this full-wave simulation. It is necessary to note that the material parameters of the new broadband non-Euclidean cloak [20] can be obtained using the above procedure (transformation optics in bipolar coordinates) apart from some factors introduced by the non-Euclidean geometry. Much effort should be put to study the properties (especially the time delay) of such a cloak. The technique proposed here shows an alternative way to implement the parameters.

## IV. Another application of the transformation optics in orthogonal coordinate

Now we come to another application of the above transformation optics when the Jacobian transformation matrix in the orthogonal coordinate has non-zero non-diagonal components. For simplicity, we suppose $\ddot{\varepsilon}_{Orthogonal} = \ddot{\mu}_{Orthogonal} = \ddot{I}$, and for the circular cylindrical coordinate, the mapping is, $(r, \theta, z) \Leftrightarrow (r', \theta', z')$,

$$\begin{cases} r' = r'(r,\theta), \\ \theta' = \theta'(r,\theta), \\ z' = z, \end{cases} \text{ or } \begin{cases} r = r(r',\theta'), \\ \theta = \theta(r',\theta'), \\ z = z', \end{cases} \tag{31}$$

So that the Jacobian transformation matrix in the circular cylindrical coordinate is,

$$T = \begin{bmatrix} \dfrac{\partial r'}{\partial r} & \dfrac{\partial r'}{r \partial \theta} & 0 \\ \dfrac{r' \partial \theta'}{\partial r} & \dfrac{r' \partial \theta'}{r \partial \theta} & 0 \\ 0 & 0 & 1 \end{bmatrix}. \tag{32}$$

with non-zero non-diagonal components in it.

(a) The rotation cloak

Chen and Chan [4] has proposed an interesting transformation media devices, the "rotation cloak" based on the following rotation mapping,

$$\begin{cases} r' = r, \\ \theta' = \theta + \theta_0 \dfrac{f(b) - f(r)}{f(b) - f(a)}, \\ z' = z, \end{cases} \tag{33}$$

It is very laborious to obtain the required parameters from the transformation optics in Cartesian coordinate [4]. However, using the above transformation optics in orthogonal coordinate, we can easily obtain the required parameters. Substitute the rotation mapping Eq. (33) into Eq. (32) to obtain the Jacobian transformation matrix in the circular cylindrical coordinate,

$$T = \begin{bmatrix} 1 & 0 & 0 \\ -\dfrac{\theta_0 r' f'(r')}{f(b) - f(a)} & 1 & 0 \\ 0 & 0 & 1 \end{bmatrix} = \begin{bmatrix} 1 & 0 & 0 \\ -t & 1 & 0 \\ 0 & 0 & 1 \end{bmatrix}. \tag{34}$$

from Eq. (7) or Eq. (10), we have,

$$\ddot{\varepsilon}_{Orthogonal}' = \ddot{\mu}_{Orthogonal}' = TT^T / \det(T)$$
$$= \begin{bmatrix} 1 & 0 & 0 \\ -t & 1 & 0 \\ 0 & 0 & 1 \end{bmatrix} \begin{bmatrix} 1 & -t & 0 \\ 0 & 1 & 0 \\ 0 & 0 & 1 \end{bmatrix} = \begin{bmatrix} 1 & -t & 0 \\ -t & 1+t^2 & 0 \\ 0 & 0 & 1 \end{bmatrix}. \tag{35}$$

Rewrite the above tensors in Cartesian coordinate, we have

$$\ddot{\varepsilon}_{Cartesian}' = \ddot{\mu}_{Cartesian}' = R \begin{bmatrix} 1 & -t & 0 \\ -t & 1+t^2 & 0 \\ 0 & 0 & 1 \end{bmatrix} R^T$$
$$= \begin{bmatrix} \cos\theta' & -\sin\theta' & 0 \\ \sin\theta' & \cos\theta' & 0 \\ 0 & 0 & 1 \end{bmatrix} \begin{bmatrix} 1 & -t & 0 \\ -t & 1+t^2 & 0 \\ 0 & 0 & 1 \end{bmatrix} \begin{bmatrix} \cos\theta' & \sin\theta' & 0 \\ -\sin\theta' & \cos\theta' & 0 \\ 0 & 0 & 1 \end{bmatrix}, \tag{36}$$

which is exactly the Eq. (5) in Ref. [4].

## (b) PEC reshaper

As the second example in this section, we review anther recent transformation media devices, the "PEC reshaper" [11] from the following mapping,

$$\begin{cases} r'(r,\theta) = \dfrac{b-a}{b-c}(r-b)+b = \dfrac{b-a}{b-c}r + \dfrac{a-c}{b-c}b, \\ \quad \theta' = \theta, \quad 0 \leq \theta < 2\pi \\ \quad z' = z, \quad z \in \mathbb{R} \end{cases}, \qquad (37)$$

where $a = \rho_1(\theta), b = \rho_2(\theta), c = \rho_3(\theta)$, which are three functions that specify the inner and outer boundaries of the transformation media and the boundary of an imaginary cylinder, respectively.

The Jacobian transformation matrix in the circular cylindrical coordinate is,

$$T = \begin{bmatrix} \dfrac{\partial r'}{\partial r} & \dfrac{\partial r'}{r\partial \theta} & 0 \\ 0 & \dfrac{r'}{r} & 0 \\ 0 & 0 & 1 \end{bmatrix} = \begin{bmatrix} \dfrac{b-a}{b-c} & \dfrac{\partial r'}{r\partial \theta} & 0 \\ 0 & \dfrac{r'}{r} & 0 \\ 0 & 0 & 1 \end{bmatrix}, \qquad (38)$$

with $\dfrac{\partial r'}{r\partial \theta} = \dfrac{a'(c-b)+b'(a-c)+c'(b-a)}{r(b-c)}\dfrac{r'-b}{b-a} + \dfrac{a-c}{b-c}\dfrac{b'}{r}$, and $\rho' = \dfrac{\partial \rho}{\partial \theta}, \rho = a,b,c,$

$r = \dfrac{b-c}{b-a}r' + \dfrac{c-a}{b-a}b$, and $\theta = \theta'$.

So, the permittivity and permeability tensors of the PEC reshaper are,

$$\ddot{\varepsilon}_{Orthogonal}' = \ddot{\mu}_{Orthogonal}' = TT^T / \det(T) = \begin{bmatrix} \mu_{rr} & \mu_{r\theta} & 0 \\ \mu_{\theta r} & \mu_{\theta\theta} & 0 \\ 0 & 0 & \varepsilon_{zz} \end{bmatrix}$$

$$= \begin{bmatrix} \dfrac{(\dfrac{\partial r'}{\partial r})^2 + (\dfrac{\partial r'}{r\partial\theta})^2}{\dfrac{\partial r'}{\partial r}\dfrac{r'}{r}} & \dfrac{\dfrac{\partial r'}{r\partial\theta}}{\dfrac{\partial r'}{\partial r}} & 0 \\ \dfrac{\dfrac{\partial r'}{r\partial\theta}}{\dfrac{\partial r'}{\partial r}} & \dfrac{\dfrac{r'}{r}}{\dfrac{\partial r'}{\partial r}} & 0 \\ 0 & 0 & \dfrac{1}{\dfrac{\partial r'}{\partial r}\dfrac{r'}{r}} \end{bmatrix}.$$

(39)

Consider the TE mode, we have, $\varepsilon_{zz} = \dfrac{1}{\dfrac{\partial r'}{\partial r}\dfrac{r'}{r}}$, and

$$\begin{bmatrix} \mu_{xx} & \mu_{xy} \\ \mu_{xy} & \mu_{yy} \end{bmatrix} = \begin{bmatrix} \cos\theta' & -\sin\theta' \\ \sin\theta' & \cos\theta' \end{bmatrix} \begin{bmatrix} \mu_{rr} & \mu_{r\theta} \\ \mu_{r\theta} & \mu_{\theta\theta} \end{bmatrix} \begin{bmatrix} \cos\theta' & \sin\theta' \\ -\sin\theta' & \cos\theta' \end{bmatrix}$$
$$= \begin{bmatrix} \mu_{rr}\cos^2\theta' - 2\mu_{r\theta}\sin\theta'\cos\theta' + \mu_{\theta\theta}\sin^2\theta' & (\mu_{rr}-\mu_{\theta\theta})\sin\theta'\cos\theta' + \mu_{r\theta}(\cos^2\theta'-\sin^2\theta') \\ (\mu_{rr}-\mu_{\theta\theta})\sin\theta'\cos\theta' + \mu_{r\theta}(\cos^2\theta'-\sin^2\theta') & \mu_{rr}\sin^2\theta' + 2\mu_{r\theta}\sin\theta'\cos\theta' + \mu_{\theta\theta}\cos^2\theta' \end{bmatrix},$$

(40)

in Cartesian coordinate. We note that the above parameters are also valid for cloak, concentrator and superscatterer with arbitrary cross section.

To show the flexibility of the present method and be different from the example given in [11], we choose $a = \rho_1(\theta) = 1\,m$, $b = \rho_2(\theta) = 2\,m$, and $c = \rho_3(\theta) = \dfrac{1\,m}{1-\cos\theta}$, where $c = \rho_3(\theta)$ is an open boundary (a parabola). A plane wave with unit amplitude and its frequency 0.1 GHz is normal incident from left to right. Fig. 3(a) and (b) plot the total electric and scattering field caused by a PEC cylinder with its outer boundary depicted by $c = \rho_3(\theta)$, respectively. Fig. 3(c) and (d) show the total electric and scattering field induced by the PEC reshaper which is a concentric cylindrical shell. By comparing the similar field patterns of Fig. 3 (a) and (c) (or (b) and (d)), we find that the concentric cylindrical shell can reshape its inner PEC core into a large parabolic PEC cylindrical. The patterns in the backward direction are almost the same while the patterns behind the scatterers have small differences due to the limited mesh elements in simulations. The large overvalued fields excited by the surface mode resonances in Fig. 3(c) and (d) are replaced with white flecks [11]. In principle, one can use these small devices as long waveguides and to guide light for a much farther distance in future designs.

(c) Kissing cloak

Finally, if we let $a = \rho_1(\theta), b = \rho_2(\theta), c = 0$, the corresponding transformation media is a cylindrical cloak with arbitrary cross section [10]. Here we set $a = 2g\cos\theta, b = 2h\cos\theta, c = 0$, the cloak should be called as "the kissing cloak" for the inner boundary and the outer boundary share the same tangent at the origin. It is similar to the structure of the non-Euclidean cloak [20]. However, the kissing cloak has singularity point at the origin while the non-Euclidean cloak does not.

To be more concrete, let $g = 0.05m$ and $h = 0.1m$, and the frequency of the incident plane wave be 2 GHz. Fig. 4(a) and (b) show the scattering pattern of the kissing cloak interacting with an incident plane wave from left to right and from bottom to top respectively. The perfect cloaking effect can be seen visually from this full-wave simulation. However there is still some small scattering from the simulation and this is due to the lack of number of the field elements and could be improved by increasing the number of the elements in principle.

## V. Discussion and conclusion

In conclusion, the transformation optics in orthogonal coordinates was proposed. The method can obtain the material parameters of the transformation media when the mapping is represented by the orthogonal coordinates. Two applications of the transformation optics were detailed with several examples, such as the elliptic cylindrical cloak, the bipolar cylindrical cloak, the rotation cloak, the PEC reshaper and the kissing cloak, *etc.*

## Acknowledgements:

I would thank Miss Xiaohe Zhang, Prof. Xudong Luo, Prof. Hongru Ma and Prof. C. T. Chan for their useful discussions. This work was supported by Hong Kong Central Allocation Fund HKUST3/06C, and the National Natural Science Foundation of China under grand No.10334020 and in part by the National Minister of Education Program for Changjiang Scholars and Innovative Research Team in University. The computation resources were also supported by the Shun Hing Education and Charity Fund.

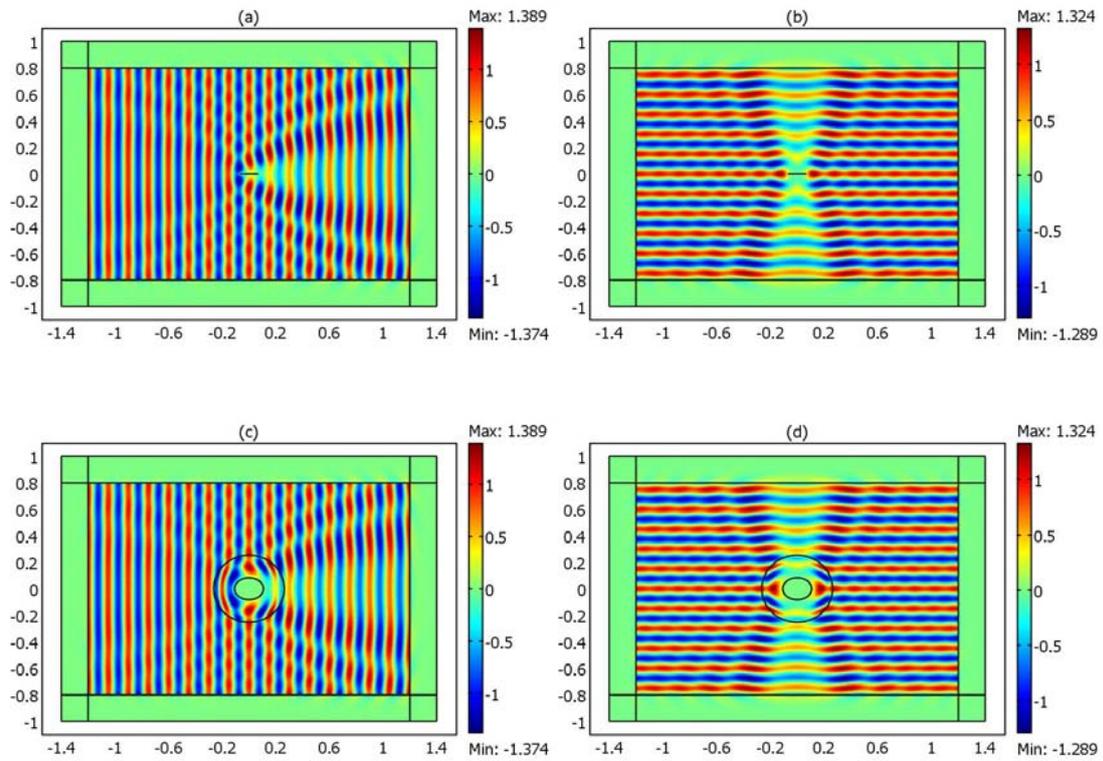

Fig. 1. (a) The scattering pattern of a PEC plate interacting with a plane wave incident from left to right. (b) The scattering pattern of a PEC plate interacting with a plane wave incident from bottom to top. (c) The scattering pattern of a PEC cylinder coating with the elliptic cylindrical cloak, and the plane wave is incident from left to right. (d) The scattering pattern of a PEC cylinder coating with the elliptic cylindrical cloak, and the plane wave is incident from bottom to top.

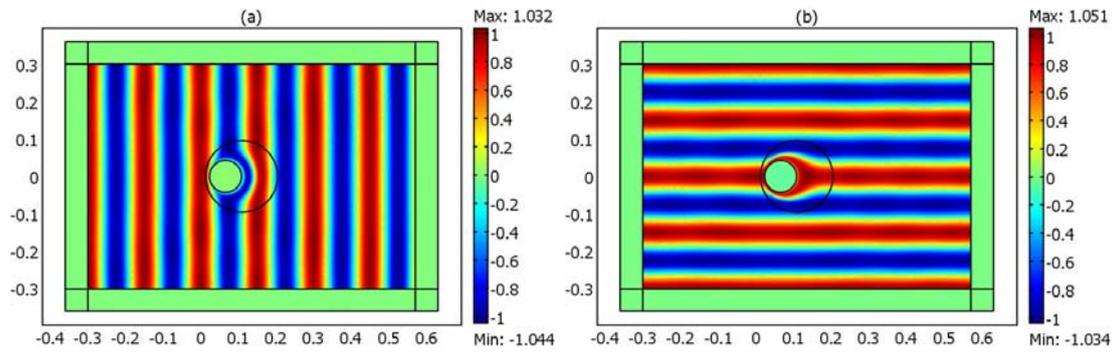

Fig. 2. (a) The scattering pattern of the bipolar cylindrical cloak interacting with a plane wave incident from left to right. (b) The scattering pattern of the bipolar cylindrical cloak interacting with a plane wave incident from bottom to top. The inner boundary of the bipolar cylindrical cloak is PEC boundary for instance.

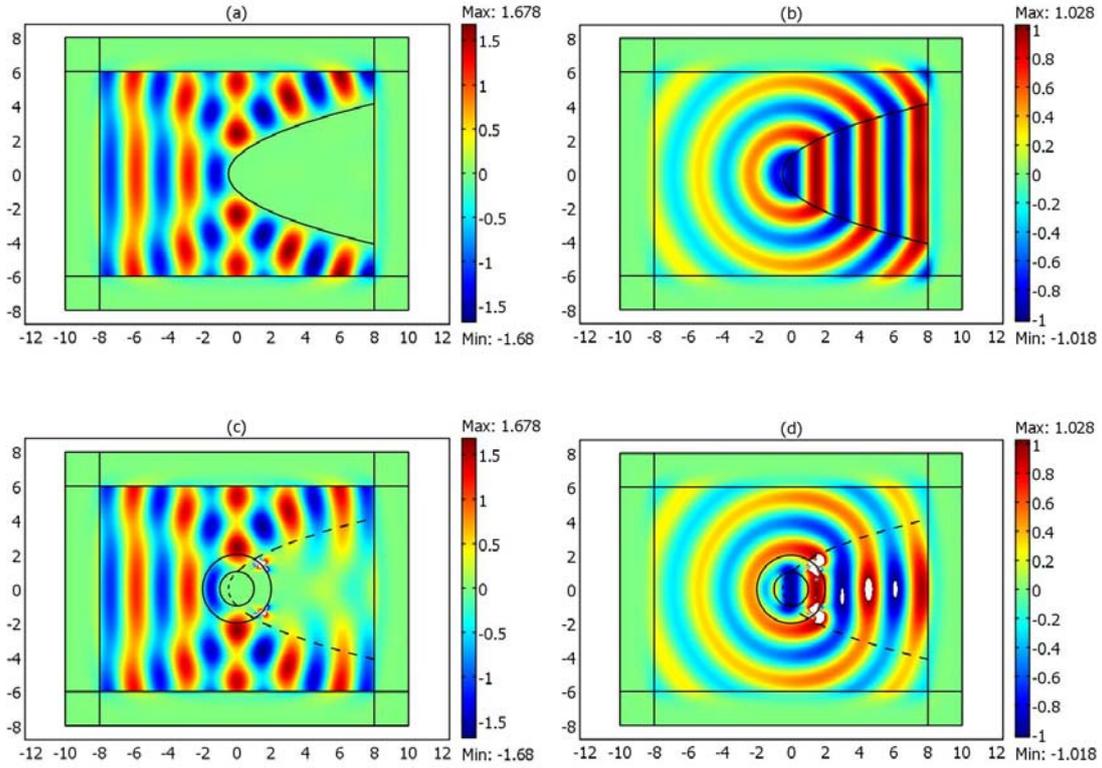

Fig. 3. (a) The total electric field of a parabolic PEC cylinder with its outer boundary depicted by $c = \rho_3(\theta)$. (b) The scattering electric field of the parabolic PEC cylinder in (a). (c) The total electric field of the designed PEC reshaper. (d) The scattering electric field of the designed PEC reshaper in (c). The dashed lines in (c) and (d) outline the boundary of the effective PEC cylinder, which is the same as the outer boundary of the PEC cylinder in (a) and (b).

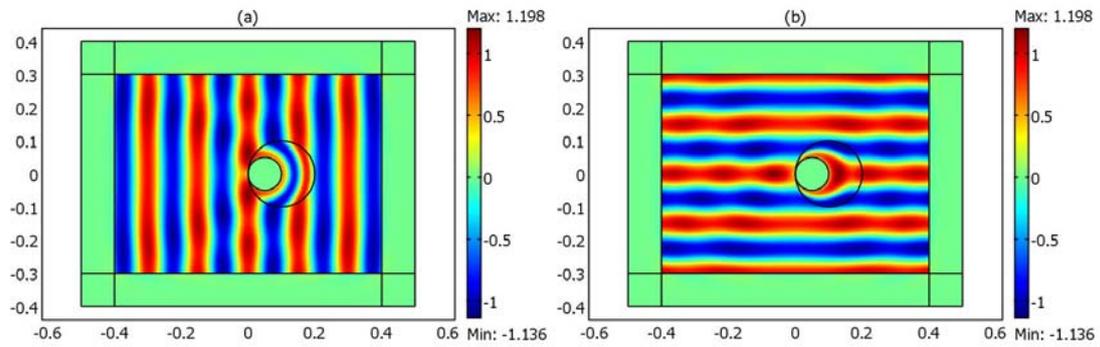

Fig. 4. (a) The scattering pattern of the kissing cloak interacting with a plane wave incident from left to right. (b) The scattering pattern of the kissing cloak interacting with a plane wave incident from bottom to top. The inner boundary of the bipolar cylindrical cloak is PEC boundary for instance.